\documentclass[apj]{emulateapj}
\usepackage{color}
\usepackage{epsfig}

\shorttitle{Electron Acceleration at Parallel Shocks}
\shortauthors{Guo and Giacalone}

\doublespace
\begin{document}

\title{The Acceleration of Electrons at Collisionless Shocks Moving Through a Turbulent Magnetic Field}

\author{Fan Guo\altaffilmark{1} and Joe Giacalone\altaffilmark{2}}

\altaffiltext{1}{Theoretical Division, Los Alamos National Laboratory, Los Alamos, NM 87545}

\altaffiltext{2}{Department of Planetary Sciences and Lunar and Planetary
Laboratory, University of Arizona, 1629 E. University Blvd., Tucson, AZ 85721}

\email{guofan.ustc@gmail.com}

\begin{abstract}
We perform a numerical-simulation study of the acceleration
of electrons at shocks that propagate through a prespecified, kinematically defined turbulent magnetic field.  The turbulence consists of broadband magnetic fluctuations
that are embedded in the plasma and cover a range of wavelengths,
the smallest of which is larger than the gyroadii of electrons that are initially
injected into the system.  We find that when the variance of the turbulent
component of the upstream magnetic field is sufficiently large -- $\sigma^2 \sim$ 10 
$B_0^2$,
where $B_0$ is the strength of the background magnetic field -- electrons can be efficiently
accelerated at a collisionless shock regardless of the orientation of the mean upstream magnetic field relative to the shock-normal direction.
Since the local angle between the incident magnetic-field vector and the shock-normal vector can be quite 
large, electrons can be accelerated through shock-drift acceleration at the shock front. 
In the upstream region, electrons are mirrored back 
to the shock front leading to multiple shock encounters. Eventually the  accelerated electrons are energetic enough that their gyroradii are of
the same order as the wavelength of waves that are included in our description
of the turbulent magnetic field. Our results are consistent 
with recent \textit{in situ} observations at Saturn's bow shock. The 
study may help understand the acceleration of electrons at shocks
in space and astrophysical systems. 
\end{abstract}


\section{Introduction}

Collisionless shocks in space and other astrophysical systems 
have been observed to be strong sources of energetic charged particles 
\citep{Blandford1987}. Diffusive shock acceleration 
\citep[DSA;][]{Krymsky1977,Axford1977,Bell1978,Blandford1978} is the
primary theory that describes quantitatively the acceleration process 
in the vicinity of shock waves. The process is expected to occur 
in many places such as propagating interplanetary shocks, planetary bow 
shocks, the solar wind termination shock, supernova blast waves, and 
shocks driven by jets from active galactic nuclei. The
theory of DSA works regardless the angle between the incident magnetic-field vector and the shock-normal vector $\theta_{Bn}$, although the 
angle can greatly influence the acceleration of charged particles 
\citep{Jokipii1982,Jokipii1987}. 
Most attention so far has been focused on the acceleration of ions. 
When the speeds of ions are large enough (usually a few times of 
the shock speed), 
they can interact resonantly with ambient magnetic turbulence and/or 
ion-scale waves and therefore get accelerated through DSA. 

The acceleration of electrons at collisionless shock waves is more poorly 
understood than that of ions. For low-energy electrons, their gyroradii are too
small for them to resonantly interact with pre-existing magnetic 
fluctuations or ion-generated waves in the shock region. The acceleration 
and scattering of low-rigidity particles, especially electrons, is thought 
to be difficult. This is usually referred to as the injection problem. For 
highly-relativistic electrons, their gyroradii are close to gyroradii of ions at the same energy 
so that the acceleration of high-energy electrons by DSA is not a problem.

Electron acceleration at quasi-perpendicular shocks ($45^\circ<\theta_{Bn}<90^\circ$) 
has been considered by a number of authors 
\citep{Wu1984,Krauss-Varban1989a,Burgess2006,Guo2010a,Guo2012b,Guo2012c}.
Analytical theories and numerical simulations have shown that at quasi-perpendicular
shocks, electrons can be accelerated through shock-drift acceleration 
\citep{Wu1984,Krauss-Varban1989a,Yuan2008,Park2012}. In this mechanism, charged
particles drift because of the gradient in the magnetic
field at the shock front. During the drift motion electrons gain
energy along the motional electric field 
$\textbf{E} = -\textbf{V} \times \textbf{B}/c$. However, it has been shown that in the
scatter-free limit, the maximum attainable energy and the fraction of particles that gain a significant amount of energy are 
very limited at a planar shock \citep[e.g.,][]{Ball2001}. Recent numerical simulations have shown 
that non-planar effects such as small-scale ripples and magnetic fluctuations
can greatly enhance the acceleration of electrons 
\citep{Burgess2006,Umeda2009,Guo2010a,Guo2012b,Yang2012}. Whistler waves may play an 
important role during the acceleration process. Observational evidence of whistler waves for the acceleration of electrons at quasi-perpendicular 
shocks has been presented \citep{Shimada1999,Oka2006,Wilson2012}.
However, the exact generation mechanism is under debate
\citep{Wu1983,Krasnoselskikh2002,Matsukiyo2006,Hellinger2007} and full
particle-in-cell (PIC) simulations performed on this subject are
limited to use an
unrealistic set of parameters including the mass ratio $m_i/m_e$ and the
ratio between Alfven speed and the speed of light $v_A/c$. For the cases with high Alfven Mach numbers and unrealistic mass ratios, some PIC simulations show that electrons can be efficiently
accelerated in the electric field due to the Buneman instability excited at the shock foot
\citep{Shimada2000,Hoshino2002,Amano2007}. However, whether this 
mechanism is robust for the realistic mass ratio $m_i/m_e = 1836$ 
in a proton-electron plasma and three-dimensional simulations is 
not clear \citep[see][]{Riquelme2011}.

The acceleration of electrons at the quasi-parallel shocks
($0^\circ<\theta_{Bn}<45^\circ$) has been less understood. 
The role of whistler waves generated 
at quasi-parallel geometry has been considered by several authors 
\citep{Levinson1992,Levinson1994,Amano2010}. In this mechanism, the thermal 
or shock-reflected electrons can generate whistler waves which 
in turn scatter electrons in pitch-angle. This mechanism requires 
a high Mach number in order for the efficient
generation of 
whistler waves \citep{Levinson1992,Amano2010}. Significant electron
acceleration has not been found in recent PIC simulations for collisonless
shocks with high mach numbers \citep{Kato2010,Riquelme2011,Niemiec2012}. 
Therefore it is not clear how electrons get efficient nonthermal 
acceleration at quasi-parallel shocks.

Effects of large-scale magnetic fluctuations have been
shown to be important for accelerating both ions and electrons at shocks \citep{Giacalone1996,Giacalone2005a,Giacalone2005b,Jokipii2007,Guo2010a,Guo2012b,Guo2012c}. 
Numerical simulations that consider large-scale 
pre-existing magnetic turbulence suggest the 
acceleration of low-rigidity particles is efficient 
and there is \textit{no} injection problem \citep{Giacalone2005a,Giacalone2005b}. 
Using self-consistent hybrid simulations 
(kinetic ions and fluid electrons) combined with 
test-particle simulations for electrons, 
\citet{Guo2010a} have found efficient electron
acceleration at perpendicular shocks moving through 
a plasma containing large-scale pre-existing magnetic 
turbulence. The turbulent magnetic field leads to 
field-line meandering that allows the electrons to 
get accelerated at the shock front multiple times. 
Small-scale shock ripples can also play a role in
scattering electrons in pitch angles similar to what is 
shown by \citet{Burgess2006}. In a more recent paper, 
\citet{Guo2012b} demonstrated that perpendicular shocks -- which exist
in some flare models -- can efficiently accelerate both electrons 
and ions. 

In previous works, wave variances of
magnetic turbulence are usually taken to be  
$\sigma^2 \leq B_0^2$ ($B_0$ is the background 
magnetic field), consistent with typical values of the variance in the interplanetary
magnetic field observed \textit{in situ} by spacecraft.
The acceleration
of electrons 
is found to prefer perpendicular shocks and there is 
no significant acceleration at quasi-parallel shocks.  
It should be noted that when there are large-amplitude ambient magnetic 
fluctuations or ion-generated waves in the upstream region, locally 
the shock angle $\theta_{Bn}$ can get quite large 
even when the shock geometry is quasi-parallel, on average, and particles
undergo acceleration by drifting along the shock due to the compressed
transverse components of the magnetic field, as in shock-drift acceleration
\citep{Guo2013a}.

\textit{In situ} observations in the heliosphere have extensive evidence
of energetic electrons associated with collisionless shocks \citep[e.g.,][]{Fan1964,Anderson1979,Gosling1989a,Simnett2005,Decker2005}. 
Examples include interplanetary shocks, planetary bow shocks, and the solar 
wind termination shock. These studies commonly found that electrons are 
accelerated at quasi-perpendicular shocks and there are rarely 
accelerated electrons 
at quasi-parallel shocks, suggesting that for those shocks,
significant electron acceleration can only occur in quasi-perpendicular
regions. Remote imaging and radio observations inferred that electrons
are accelerated at quasi-perpendicular regions of coronal shocks \citep{Kozarev2011,Feng2012,Feng2013}. This indicates that for solar
energetic particle (SEP) events, where the
energetic electrons are observed to have tight correlations with
energetic ions, quasi-perpendicular shocks may accelerate most
energetic particles \citep{Haggerty2009,Cliver2009,Guo2012c}. 
However, in astrophysical shocks such as supernova blast waves, observations suggest that electrons can be accelerated to highly 
relativistic energy regardless of the shock angle, $\theta_{Bn}$, although the local shock angle is not directly observable, and, therefore is not known
\citep{Reynolds2008}. For example, Chandra's 
observations for Tycho supernova remnant have revealed strong non-thermal 
X-ray emissions surrounding the remnant, presumably caused by synchrotron radiations of strongly accelerated electrons in the shock region \citep[e.g.,][]{Eriksen2011}. 

Recently, \citet{Masters2013} 
reported a rare \textit{in situ} measurement where electrons can be 
accelerated to $\sim MeV$ in the quasi-parallel part of Saturn's bow shock. 
The shock is observed to be associated with large-amplitude magnetic 
fluctuations ($\delta B/B_0 \sim 10$ or even larger). This indicates electrons can be efficiently accelerated at quasi-parallel shocks when large-amplitude magnetic fluctuations are present in the shock region.

In this paper, we explore the effect of strong large-scale magnetic fluctuations
on the acceleration of electrons at collisionless shocks. We show electrons 
can be  efficiently accelerated to high energy regardless of the 
angle between the average magnetic field and shock normal 
when strong large-scale magnetic fluctuations exist in the shock region 
($\sigma^2 \sim 10 B_0^2$). 
This finding can help understand the acceleration of electrons at astrophysical shocks such as recent \textit{in situ} observation at Saturn's bow shock \citep{Masters2013} and radio and X-ray observations for supernova shocks \citep{Reynolds2008}.

\section{Numerical Method}

We integrate numerically the equations of motion of an ensemble 
of electrons in pre-specified, kinematically defined 
electric and magnetic fields in the 
shock region. This method resolves the gyromotions of 
charged particles close to the shock that is critical to the 
acceleration of low-rigidity particles.
The electrons are assumed to have a negligible
effect on the shock fields; thus, they are treated as test particles.  The approach is similar to previous studies \citep{Decker1986,Decker1988,Giacalone1996,Giacalone2005a} and we use it for electrons as was done by \citet{Giacalone2009}. Here we describe the salient details of the numerical method for completeness, the readers are referred to the previous papers for further details. 

In the simulation, a planar shock is located at $x = 0$, and plasma flows from $x < 0$ (upstream) with a speed of $U_1$ to $x > 0$ (downstream) with a speed of $U_2 = U_1/r$, where $r$ is the compression ratio. The flow speed along the $x$ direction is assumed to be
\begin{eqnarray}
U(x) = \frac{2U_1}{(r+1)+(r-1)\tanh(x/\delta_{sh})},
\end{eqnarray}

\noindent where $\delta_{sh}$ is the width of the shock. The upstream speed in the shock frame is taken to be $U_1 = 500$ km/s. Close to the shock, the flow speed varies smoothly across a sharp layer with $\delta_{sh} = 0.2 U_1/\Omega_i$, where $\Omega_i$ is the upstream proton cyclotron frequency. This is similar to the observed high-Mach-number shocks \citep{Bale2003}. We assume a compression ratio for the strong shock limit $r = U_1/U_2 = 4$, corresponding to the case where both the upstream Alfvenic Mach number $M_A$ and sonic Mach number $M_s$ are much larger than 1. The dynamical evolution of the shock surface caused by the convection of magnetic fluctuations across the shock is neglected since the dynamical pressure is much larger than the magnetic pressure.

The magnetic field embedded in the plasma flow is given by the solution to the magnetic induction equation. In our model, the magnetic-field components are given by the convected magnetic field
\begin{eqnarray}
&B_x(x, y, z, t)& = B_x(x_0, y, z, t_0), \nonumber \\
&B_y(x, y, z, t)& = B_y(x_0, y, z, t_0) \left(\frac{U(x_0)}{U(x)} \right),  \\
&B_z(x, y, z, t)& = B_z(x_0, y, z, t_0) \left(\frac{U(x_0)}{U(x)} \right), \nonumber
\end{eqnarray}

\noindent where $x_0$ and $t_0$ can be related using the equation $dx/dt = U(x)$. Note that the transverse components of magnetic fields increase when the plasma is compressed. The solution to the equation can be expressed as
\begin{eqnarray}
t - t_0 = \int^x_{x_0} \frac{dx'}{U(x')}.
\end{eqnarray}

We choose $x_0$ to be far upstream of the shock, and use Equation 
(3) to determine $t_0$ using the speed of plasma flow given in 
Equation (1).  Using these values in Equation (2), the magnetic 
field vector is defined at any point in space and time as long as 
it is known at ($x_0, y, z, t_0$). The field at this point is 
taken to be $\textbf{B} (x_0, y, z, t_0) = \textbf{B}_0(x_0) + 
\delta \textbf{B} (x_0, y, z, t_0)$, where $\textbf{B}_0$ is the 
average magnetic field lying in the $x$-$z$ plane and $\delta 
\textbf{B}$ is a random magnetic field component.  For the 
fluctuation component, we assume purely isotropic magnetic 
fluctuations \citep{Batchelor1953}, which is approximated by a 
large number of wave modes having wavevectors randomly 
distributed in direction and with random phases and 
polarizations. The amplitude of
each mode is determined from an assumed Kolmogorov-like power 
spectrum. For more details on generating the turbulent magnetic 
fields, see \citet{Giacalone1999}.
 In our simulations, the correlation length is
taken to be $L_c = 50 U_1/\Omega_i$. The maximum and 
minimum wavelengths used to generate the turbulence are $\lambda_{max} = 500 U_1/\Omega_i$ and 
$\lambda_{min} = 5 U_1/\Omega_i$, respectively. Under typical conditions 
near Saturn's bow shock, the coherence length corresponds to a 
spatial scale of $\sim 10^5$ km.
The motional 
electric field is obtained under the ideal MHD approximation 
$\textbf{E} = - \textbf{U} \times \textbf{B}/c$. It is important 
to note that our simulations utilize a fully three-dimensional 
magnetic field. This is essential for correctly considering 
cross-field diffusion as noted by previous works
\citep{Jokipii1993,Giacalone1994a,Jones1998}.

In the simulation the electrons are injected upstream of the 
shock  at a constant rate. This is done by initializing electrons 
immediately upstream ($x = -5 U_1/\Omega_i$) with a time randomly 
chosen between $t = 0$ and $t_{max}$.   The initial velocity 
distribution is isotropic in the local plasma frame with an 
energy of $1$ keV. The relativistic equations of motion 
\begin{eqnarray}
\frac{d \textbf{p}}{dt} = e (\textbf{E} + \textbf{v} \times \textbf{B}), \\
\frac{d \textbf{x}}{dt} = \textbf{v},
\end{eqnarray}
for each 
particle are numerically integrated using the Burlirsh-Stoer 
method \citep{Press1986}. The algorithm uses an adjustable time 
step based on an evaluation of the local truncation error. It is highly accurate and fast when fields are smooth compared with the electron gyroradius. The 
energy is conserved to an accuracy of better than $0.1\%$ of the 
total energy in the plasma frame when the shock is not considered 
in our simulation. For the parameters we use, the initial 
gyroradii of electrons are about $0.02 U_1/\Omega_i$, much less 
than the thickness of the shock layer and the spatial scale of 
the injected magnetic fluctuations. In the end of the 
simulations, the maximum energy of electrons reaches energies exceeding an MeV 
and their gyroradii become large enough for them to 
interact resonantly with the injected magnetic fluctuations.
No spatial boundary is placed in the simulation. We
numerically integrate the trajectory of about $5$ million test particles
for each case. The trajectories for
all the electrons are integrated until $\Omega_i t_{max} = 400.0$. At the end of the simulation, some accelerated electrons 
can propagate to a distance of several thousand $U_1/\Omega_i$ away 
from the shock. The electrons considered in this work are tracked in a  
much larger spatial region than that is achievable in recent two- and three-dimensional particle-in-cell simulations \citep{Giacalone2000b,Guo2010a,Guo2013a,Caprioli2013}.
We vary the turbulence variance, $\sigma^2$, and angle between 
the average magnetic field and shock normal, $\theta_{Bn}$, in 
our calculations in order to understand their effect on the 
energization of electrons, and particularly on the energy 
spectrum.
We also calculate the
fraction of electrons with energies $E > 10$ keV at the end of the simulation. 
This is used to characterize the efficiency of electron acceleration. Table \ref{table1} lists the parameters for each run.

We assume that the electrons have a negligible effect on the electric and magnetic fields close to 
the shock. We also neglect shock microstructure such as magnetic overshoot, and cross-field electric 
field, etc. Note that those effects modify the motion of a single particle at the shock front, but 
do not change the main conclusion of the current study. For example, 
considering magnetic overshoot can increase the population of electrons that gain 
energy through adiabatic reflection \citep{Wu1984}. The 
cross-shock potential is much less than the 
initial electron energy per charge, 
meaning ignoring that effect does not significantly change the results.

\begin{table}
\centering
\begin{tabular*}
{0.3\textwidth}{cccc}
\hline
Run& $\sigma^2 (B_0^2)$ & $\theta_{Bn}$ & $\Gamma\% (E>10keV)$  \\
\hline
1  & 0.1 & 0  & 0.0\\
2  & 0.1 & 15 & 0.0\\
3  & 0.1 & 30  & 0.0\\
4  & 0.1 & 45  & 0.0\\
5  & 0.1 & 60 & 0.2\\
6  & 0.1 & 75  & 0.97\\
7  & 0.1 & 90  & 9.3\\
8  & 1.0 & 0  & 1.8\\
9  & 1.0 & 15 & 2.1\\
10  & 1.0 & 30  & 2.58\\
11  & 1.0 & 45  & 3.78\\
12  & 1.0 & 60 & 6.04\\
13  & 1.0 & 75  & 9.5\\
14  & 1.0 & 90  & 13.5\\
15  & 10 & 0  & 7.3\\
16  & 10 & 15 & 7.5\\
17  & 10 & 30  & 7.9\\
18  & 10 & 45  & 8.6\\
19  & 10 & 60 & 9.7\\
20  & 10 & 75  & 10.9\\
21  & 10 & 90  & 11.4\\
 \hline
\end{tabular*}
 \caption{Parameters for each simulation run. The total wave variance $\delta B^2/B_0^2$. The averaged shock normal angle $\theta_{Bn}$, and the fraction of electrons whose energy is more than $10$ keV at the end of simulation.}
 \label{table1}
\end{table}

\section{Simulation Results}

Using numerical simulations described in Section 2, we examine the effects of wave variances $\sigma^2$ and average shock angles $\theta_{Bn}$ on the acceleration of electrons at shocks. Table \ref{table1} lists the parameters for each run. It also contains the fraction of electrons with energies $E > 10$ keV at the end of the simulation for each simulation run. This is used to characterize the efficiency of electron acceleration.

\begin{figure}
\begin{tabular}{c}
\includegraphics[width=0.45\textwidth]{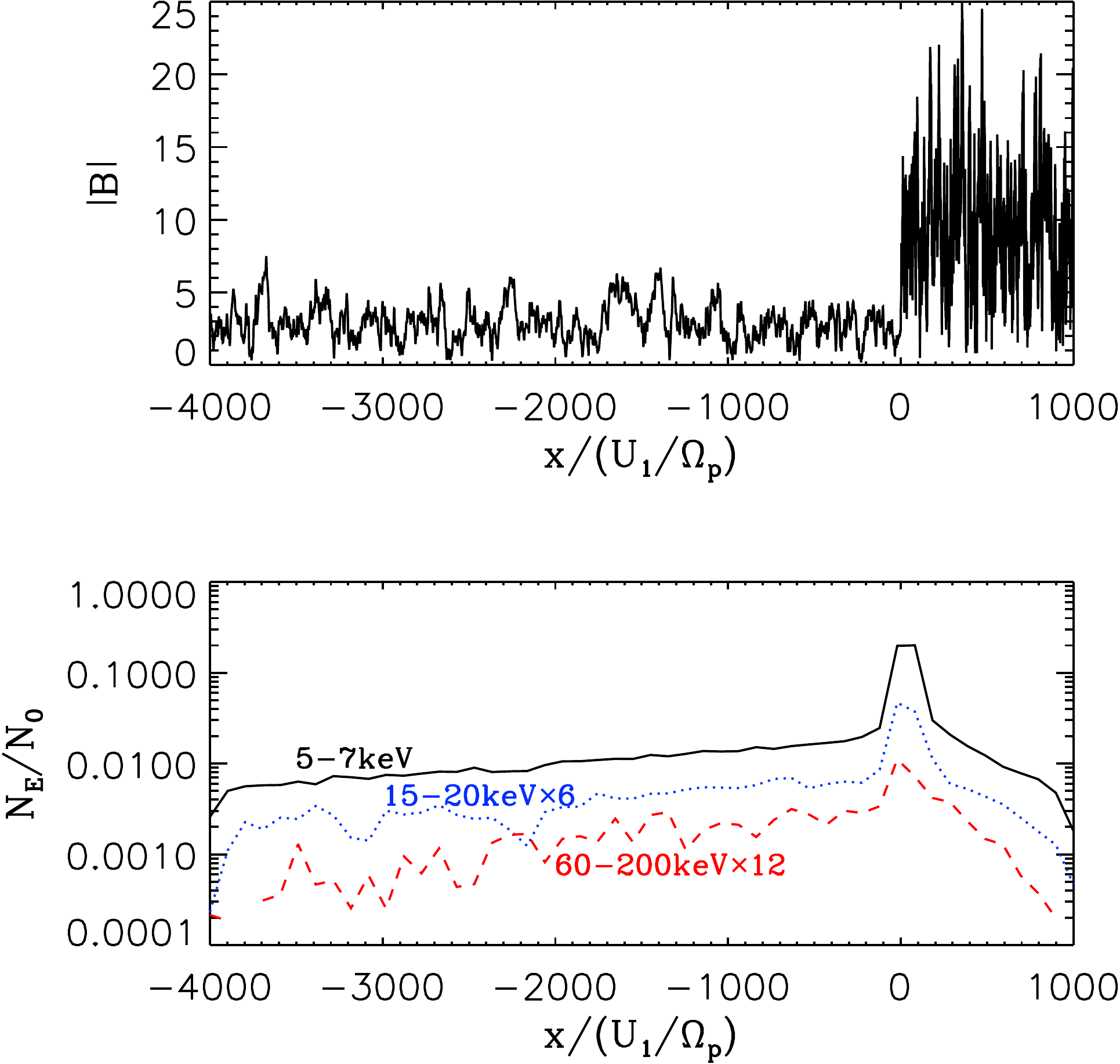}
\end{tabular}
\caption{Upper panel: The magnitude of magnetic field across the shock wave. Lower panel: the averaged density of accelerated electrons in energy ranges $5$-$7$ keV (black solid line), $15$-$20$ keV (blue dotted line), and $60$-$200$ keV (red dashed line). \label{Fig1}}
\end{figure}

Figure \ref{Fig1} shows the magnetic field magnitude along the $x$ direction in the 
upper panel. In this case (Run $16$), the average shock angle 
$\theta_{Bn} = 15^\circ$ and the wave variance of magnetic fluctuations 
$\sigma^2 = 10B_0^2$. The upstream magnetic field is featured by large-amplitude 
magnetic fluctuations. The magnitude of the magnetic-field increases at the shock 
due to the compression of its transverse components. 
The lower panel shows the averaged spatial
distributions of accelerated electrons along the $x$ direction
at $\Omega_i t=400.0$. The black solid, blue dotted, and red
dashed curves represent the density of accelerated electrons averaged over the $y$ and $z$ directions with energy ranges
$5$-$7$ keV, $15$-$20$ keV, and $60$-$200$ keV, respectively. We
find that electrons can be accelerated to relativistic energies 
at the quasi-parallel shock. The
accelerated electrons concentrate close to the shock,
indicating that electrons gain energy right at the
shock front. At the end of the simulation, the maximum energy of the accelerated electrons has reached $\sim 1$ MeV. Those electrons have gyroradii large enough for them to resonantly interact with the injected magnetic fluctuations.

The strikingly efficient electron acceleration at a 
quasi-parallel shock has not been seen in previous numerical
simulations \citep{Guo2010a}.
As large-amplitude magnetic fluctuations convect across the shock, 
locally the shock can have a quasi-perpendicular geometry, which is
expected to accelerate particles \citep{Decker1986,Giacalone2005a}. 
And, this occurs for long enough of a time interval that the electrons gain 
significant energy.
The effect 
of the magnetic variance of upstream fluctuations on 
electron acceleration at perpendicular shocks 
has been studied by \citet{Guo2010a,Guo2012b,Guo2012c}.

\begin{figure}
 \includegraphics[width=0.45\textwidth]{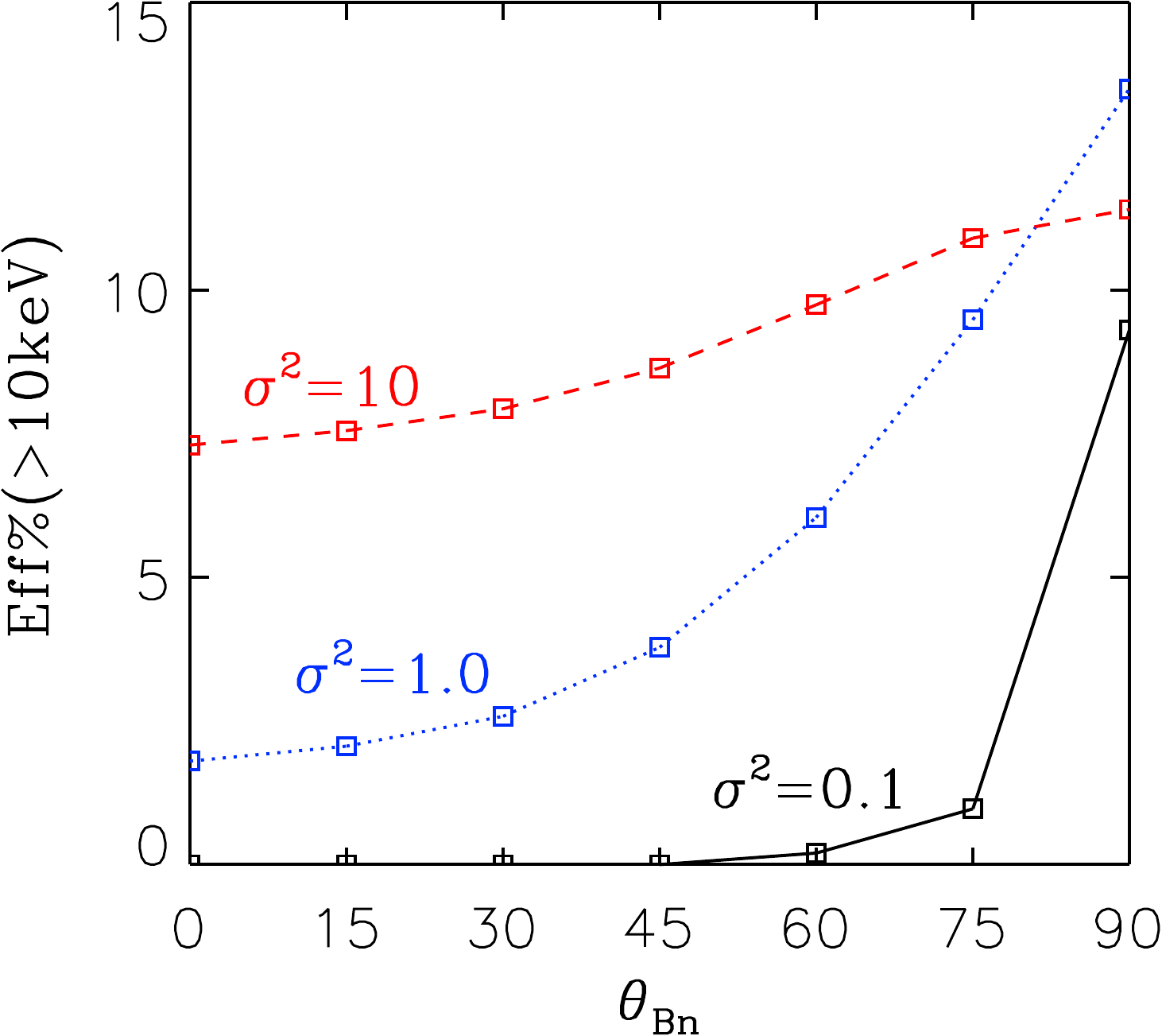}

\caption{The efficiency of electron acceleration for various wave variances as a function of shock angle. $\Gamma \%$ is defined by the fraction of electrons
that is accelerated to more than $10$ keV at the end of the simulations. \label{Fig2}}
 \end{figure}

Figure \ref{Fig2} shows $\Gamma \%$, the fraction of
accelerated electrons with $E > 10$ keV at the end of simulations as a 
function of $\theta_{Bn}$ for
various wave variances $\sigma^2 = 0.1B_0^2$ (black solid line), $1.0 B_0^2$ (blue 
dotted line), and $10.0 B_0^2$ (red dashed line). 
The fraction for each run is also listed in Table \ref{table1}. 
One can see that for 
the cases with $\sigma^2 = 0.1 B_0^2$ and $1.0 B_0^2$, the efficiency of electron
acceleration is strongly dependent on the shock normal angle. However, for
the case that $\sigma^2 = 10 B_0^2$, the accelerated fraction varies from 
$7.3 \%$ for $\theta_{Bn} = 0^\circ$ to $11.4 \%$ for 
$\theta_{Bn} = 90^\circ$. This 
shows that the acceleration of electrons depends weakly on the shock angle
when the wave variance in the upstream region is sufficiently large. Note that 
in this case, shocks with larger shock angles can still accelerate electrons more efficiently than shocks with smaller shock angles. It is also interesting to note that in the case with $\sigma^2 = B_0^2$ and $\theta_{Bn} = 90^\circ$ (Run 14), the fraction of electrons that reach $10$ keV is higher than that in the case with larger turbulence variance $\sigma^2 = 10 B_0^2$ and $\theta_{Bn} = 90^\circ$ (Run 21). This is because the local shock angles where particles interact with the shock is effectively reduced when stronger turbulence present in the shock region. For low energy particles, this effect makes more electrons accelerated to energies higher than $10$ keV for
the case with turbulence variance $\sigma^2 = B_0^2$.

\begin{figure}
\includegraphics[height=0.55\textheight]{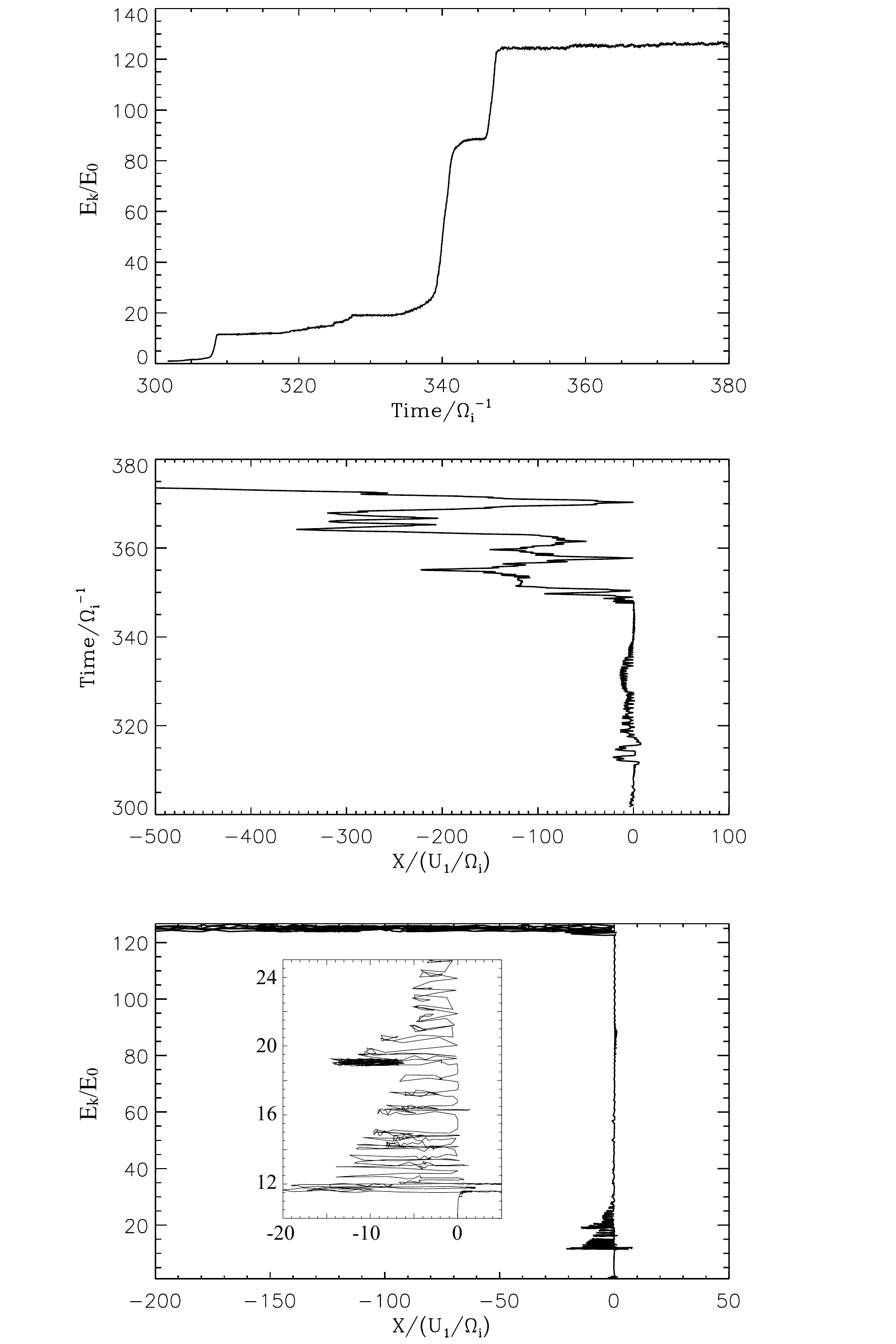}
\caption{The trajectory analysis for an accelerated electron. The upper two panels show the evolution of energy and $x$ position between $t = 120 \Omega_{i}^{-1}$ and $t = 230 \Omega_{i}^{-1}$. The middle panels show the time
evolution of particle energy and the magnitude of magnetic field at the location of the electron during period `$a$'. The bottom panels show similar quantities during period `$b$'. \label{Fig3}}
 \end{figure}

Figure \ref{Fig3} shows the trajectory of a representative electron that is
accelerated to more than $100E_0$ at a quasi-parallel shock with $\theta_{Bn} = 
15^\circ$ and $\sigma^2 = 10 B_0^2$ (Run $16$). The upper panel displays
the time evolution of the particle energy, the middle panel shows the position in the $x$ direction, and the lower panel shows the energy as a function of position
in the $x$ direction between about $t = 300 \Omega_{i}^{-1}$ (the injection time) and 
$t = 380 \Omega_{i}^{-1}$, respectively. The inset shows a blow-up close to the shock. The figure shows that the particle interacts
with the shock numerous times and by which gains a large amount of energy. There are several
large energy increases (e.g., during time period $t = 335 \Omega_{i}^{-1}$ - $345 \Omega_{i}^{-1}$)
where the electron gains energy several times of its energy before the 
shock encounter. It also shows many small accelerations during which particles keep
returning to the shock and obtaining multiple energy increases at the shock front.
In the upstream region, the electron can also gain a small amount of energy as it 
gets reflected in the upstream medium.
 Recent publications have emphasized
this process by using one-dimensional PIC simulations \citep{Kato2014,Park2014}, 
but the shock speed is 
much larger $U_1 = 0.1-0.3 c$ and the energy
gain in each upstream reflection is $\Delta E \sim 2U_1 E/c $ in the downstream frame. For nonrelativistic shocks such as heliospheric
shocks and supernova blast waves, this effect will be much reduced as we have shown here.

To better understand this, we closely examine the trajectory between $t = 300 \Omega_{i}^{-1}$ (the injection time) and 
$t = 325 \Omega_{i}^{-1}$. The upper two panels show 
the time evolutions of the particle's energy and position in the $x$ 
direction respectively. During the time period `$a$', 
the particle gains energy about $7$ times of its energy before the shock encounter.
In the middle panels, we closely examine the trajectory between 
$\Omega_i t = 306$ - $310$, which is the time period marked by 
`$a$' in the upper panels. The two plots show the time evolution of energy and the 
magnitude of magnetic field at the position of the electron. They clearly show that 
the electron gains energy while the magnetic field seen by the particle increases. 
This is typical of shock-drift acceleration. In the lower panels, we analyze the 
time period marked by `$b$' in the upper panels. In this time duration, the electron 
has multiple shock encounters and each encounter corresponds to a small energy 
increase. Although we do not include fluctuations at scales associated with
the gyroradii of the electrons at the injection energy which would cause 
pitch-angle scattering, we find that electrons can be mirrored back to the shock
when they encounter a sufficiently large magnetic field magnitude upstream of 
the shock. Note that when the electron is close to the shock, the typical spatial scale for the particle to get reflected immediately upstream is $\sim 20 U_1/\Omega_i$,
consistent with the scale of the upstream waves.

 \begin{figure}
\includegraphics[width=0.5\textwidth]{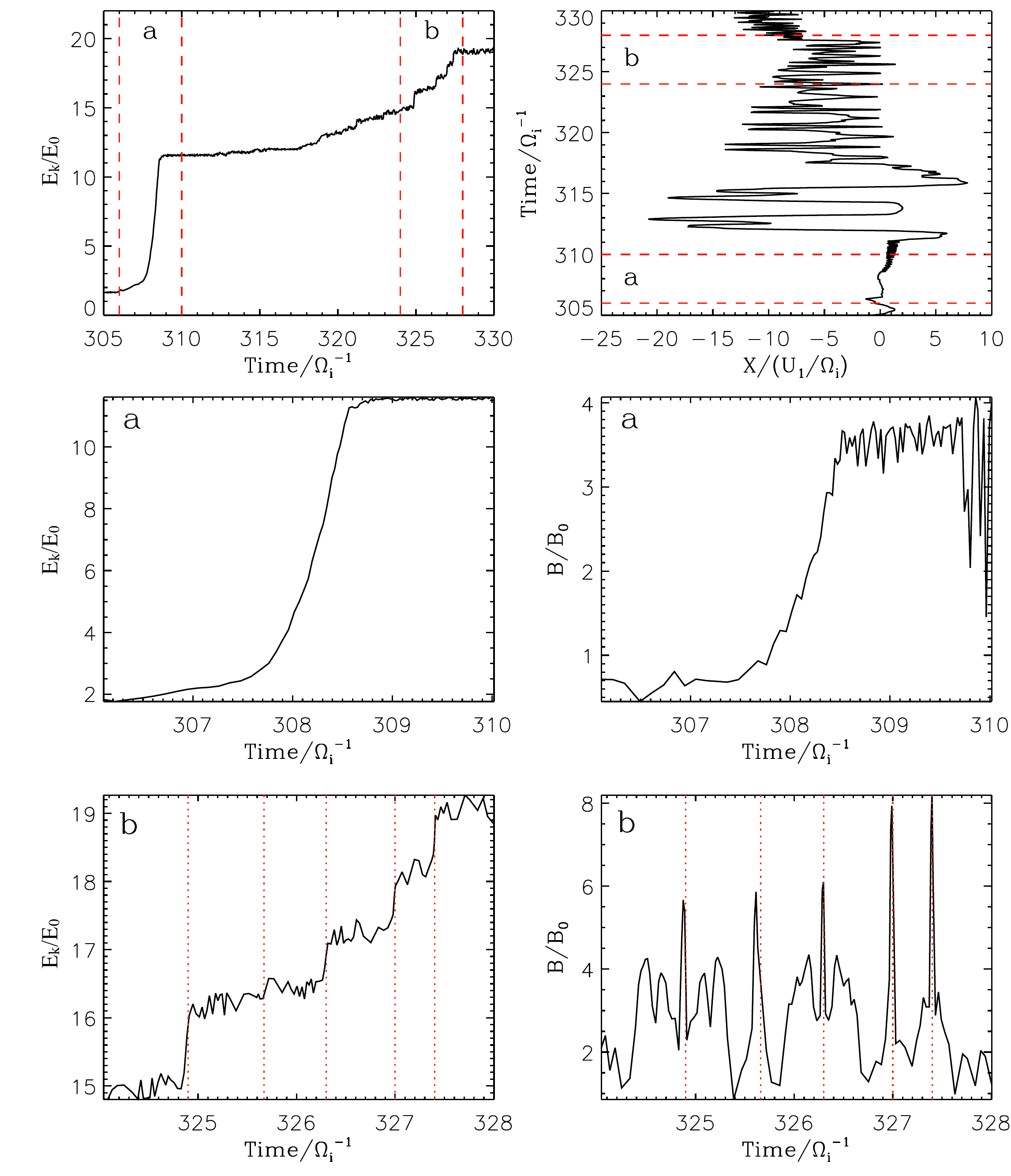}
\caption{The trajectory analysis for an accelerated electron. The upper two panels show the evolution of energy and $x$ position between $t = 120 \Omega_{i}^{-1}$ and $t = 230 \Omega_{i}^{-1}$. The middle panels show the time
evolution of particle energy and the magnitude of magnetic field at the location of the electron during period `$a$'. The bottom panels show similar quantities during period `$b$'. \label{Fig4}}
 \end{figure}

\begin{figure}
\includegraphics[width=0.45\textwidth]{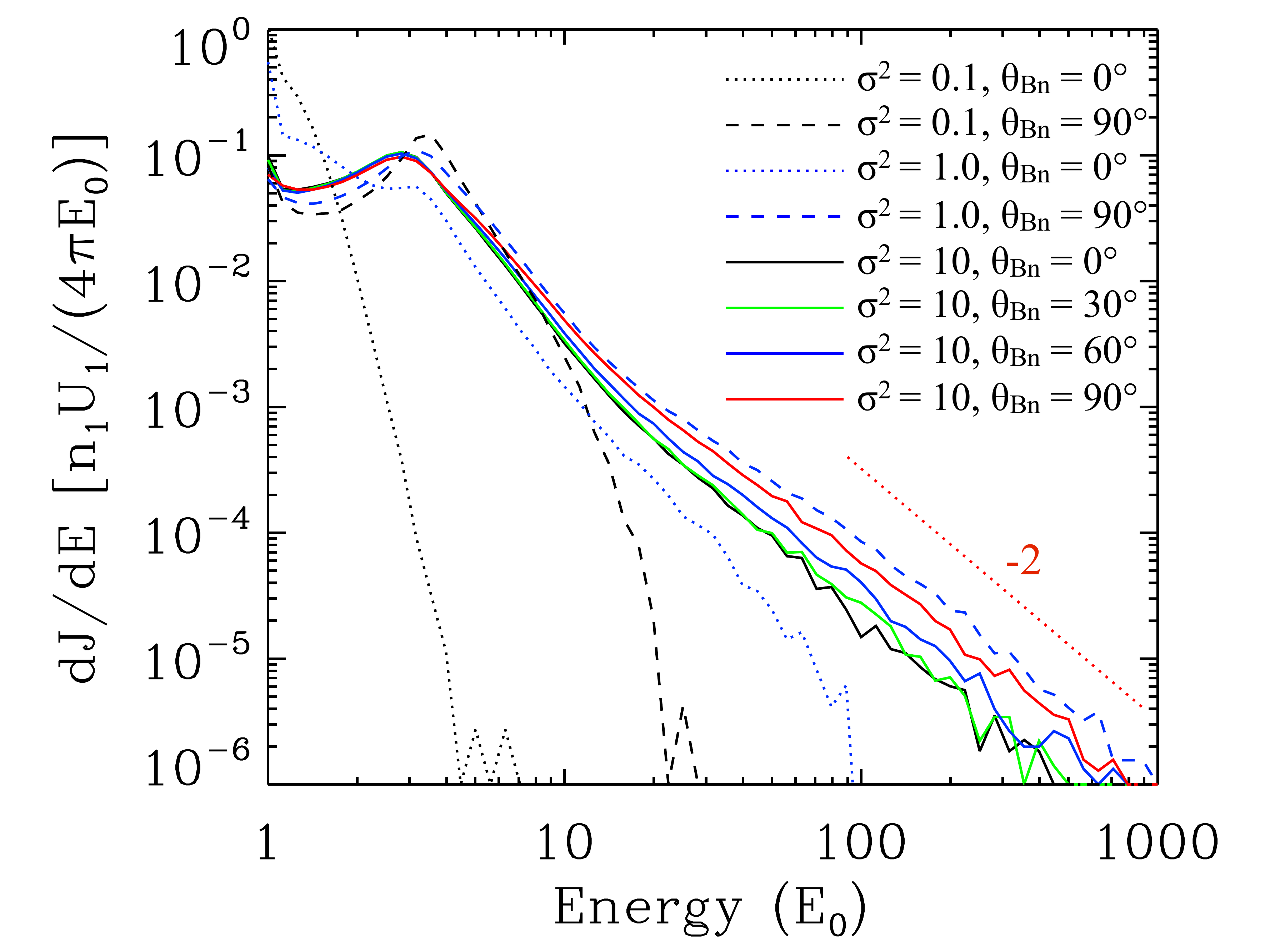}
\caption{Energy spectra of electrons in the downstream region at the end of simulation for a variety of runs. The black, blue, green and red solid lines are for $\sigma^2 = 10 B_0^2$ with $\theta_{Bn} = 0^\circ$ (Run 15), $30^\circ$ (Run 17), $60^\circ$ (Run 19), and $90^\circ$ (Run 21), respectively. For comparison, the spectra for $\sigma^2 = 0.1 B_0^2$ with $\theta_{Bn} = 0^\circ$ (Run $1$), and $90^\circ$ (Run $7$) are shown by the black dotted line and black dashed line, and the spectra for $\sigma^2 = B_0^2$ with $\theta_{Bn} = 0^\circ$ (Run $8$), and $90^\circ$ (Run $14$) are shown by the black dotted line and black dashed line, respectively.\label{Fig5}}
 \end{figure}

Figure \ref{Fig5} shows the downstream energy spectra of electrons at the end of simulation for a 
variety of runs. The black, blue, green and red solid lines are for $\sigma^2 = 10 B_0^2$ with 
$\theta_{Bn} = 0^\circ$ (Run 15), $30^\circ$ (Run 17), $60^\circ$ (Run 19), and $90^\circ$ 
(Run 21), respectively. For comparison, the spectra for $\sigma^2 = 0.1 B_0^2$ with 
$\theta_{Bn} = 0^\circ$ (Run $1$), and $90^\circ$ (Run $7$) are shown by the black dotted line and 
black dashed line, and the spectra for $\sigma^2 = B_0^2$ with 
$\theta_{Bn} = 0^\circ$ (Run $8$), and $90^\circ$ (Run $14$) are shown by the blue dotted line and 
blue dashed line, respectively.
 One can see that when the wave variance of magnetic fluctuations is sufficiently 
 large, the resulting energy spectrum does not significantly depend on the
 average shock-normal angle.
 In all the cases with $\sigma^2 = 10 B_0^2$, electrons can be accelerated up to $\sim 1$ MeV. 
 The spectra at
 quasi-parallel shocks are consistent with the recent observation of electron 
 acceleration at a high-mach-number quasi-parallel shock associated with strong 
 magnetic fluctuations \citep{Masters2013}. When the wave variance is 
 $\sigma^2 = 0.1 B_0^2$, perpendicular shocks can accelerate electrons more 
 efficiently than parallel shocks and the resulting distributions depend strongly 
 on the average shock normal angle. In this case the maximum 
 electron energy can only 
 reach $30$ keV in the perpendicular shock case and $7$ keV in the 
 parallel shock case. When the wave variance is $\sigma^2 \geq B_0^2$, the energy spectra
 of accelerated electrons do not change much at perpendicular shocks, meaning the 
 acceleration of electrons saturates when the wave variance is sufficiently large.

\section{Discussion and Conclusions}

In this paper, using numerical simulations, we calculated the trajectories of a large number of
electrons encountering a shock that moves through a strongly fluctuating magnetic field.
We found that the large-amplitude magnetic fluctuations have a significant effect on the acceleration of electrons. For the case that the
wave vaiance $\sigma^2 \le 1.0 B_0^2$, the acceleration of
electrons strongly depends on the average shock normal angle. However,
in the case that $\sigma^2 \sim 10 B_0^2$, electrons can
be accelerated efficiently to relativistic energies regardless of the
shock angle. This indicates that the acceleration of electrons 
is weakly dependent on the average shock normal angle when the 
upstream wave variance is sufficiently large. This 
is also consistent with recent observation by \citet{Masters2013}, 
who reported \textit{in situ} measurements showing that
electrons get accelerated to relativistic energies at 
a high-mach-number quasi-parallel shock that is associated with
strong magnetic fluctuations with $\delta B/B_0 \sim 10$ or larger. We find 
electrons can be reflected by strong magnetic field in the upstream region and get 
accelerated at the shock though drift acceleration. The energy spectra of electrons in 
the end of the simulation for different shock angles are remarkably similar, 
indicating that they are 
accelerated by the same process. In our simulation, electrons are 
accelerated up to $\sim 1 MeV$ within several hundred proton gyroperiods. At that energy the accelerated electrons have gyroradii large enough to resonantly interact with the injected magnetic fluctuations. 
This provides an efficient mechanism for injecting electrons into diffusive shock acceleration and is important to explain electron acceleration and high-energy emissions at astrophysical shocks \citep{Kang2012}. 

Finally, we note that although this study shows that the acceleration of 
electrons can be efficient at quasi-parallel shocks when there exists large-
amplitude magnetic fluctuations, which is consistent with the observation made by 
\citet{Masters2013}, the origin of the strong magnetic fluctuations is not clear. 
Large-amplitude magnetic fluctuations have been inferred to be 
present in the vicinity 
of high-Mach-number supernova shocks \citep{Berezhko2003}. However, the dominate 
mechanism is still under debate 
\citep{Bell2004,Giacalone2007,Amato2009,Inoue2009,Guo2012a,Greenfield2012}. The 
\citet{Masters2013} observation shows that ions are only accelerated to $\sim 10$ 
keV, indicating in this case the effect of energetic protons on generating strong 
magnetic fluctuations is not significant. Studying the generation of large-amplitude 
magnetic fluctuations is beyond the scope of the present study and we will revisit 
this problem in future work.

\section*{Acknowledgement}
F.G. benefited from discussion with Dr. Hongqing He and Dr. Yi-Hsin Liu. This work was supported by NASA under grant NNX11AO64G and by NSF under grant AGS1154223 and AGS1135432. Part of the computational resource supporting this work were provided by the institutional computing resources at Los Alamos National Laboratory.

\singlespace

\clearpage

\end{document}